\documentclass[twocolumn,showpacs,preprintnumbers,amsmath,amssymb]{revtex4}

\usepackage{graphicx}
\usepackage{dcolumn}
\usepackage{bm}

\begin{document}

\title{Electric-field induced phase transitions in rhombohedral Pb(Zn$_{1/3}$Nb$_{2/3}$)$_{1-x}$Ti$_{x}$O$_{3}$}
\author{B. Noheda}
 \altaffiliation {Corresponding author, e-mail: noheda@bnl.gov}
\author{Z. Zhong, D.E. Cox, and G. Shirane}
\affiliation{Brookhaven National Laboratory, Upton, New York 11973\\}
\author{S-E. Park}
\affiliation{Fraunhofer-IBMT Technology Center Hialeah, Hialeah, Florida 33010\\}
\author{P. Rehring}
\affiliation{Materials Research Laboratory, The Pennsylvania State University, University Park, Pennslyvania 16802\\}

\date{\today}

\begin{abstract}
High-energy x-ray diffraction experiments peformed on rhombohedral Pb(Zn$%
_{1/3}$Nb$_{2/3}$)$_{1-x}$Ti$_{x}$O$_{3}$ (PZN-x\%PT) crystals with x= 4.5
and 8\% show that an electric field applied along the [001] direction
induces the tetragonal phase, as proposed by Park and Shrout. Our
experiments reveal that in PZN-4.5\%PT such phase change occurs 
\textit{via} a third phase with monoclinic symmetry, M$_{A}$, which is
observed at intermediate field values. This is in agreement with
first-principles calculations by Fu and Cohen predicting the rotation of
the polarization between the rhombohedral and tetragonal phases in this
material. A different polarization path between the rhombohedral and
tetragonal phases, through a second monoclinic phase, M$_{C}$, has been
previously reported in PZN-8\%PT. The microscopic characterization of these crystals
allows us to explain the ultra-high macroscopic strain observed in PZN-x\%PT
under an electric field. Moreover, some unusual scattering profiles
displayed by exceptionally good crystals, are experimental evidence of the
high anharmonicities and near-degeneracy of the different phases in these
extremely deformable materials.
\end{abstract}
\maketitle


\section{Introduction}

Piezoelectric single crystals of the relaxor- ferroelectric material Pb(Zn$_{1/3}$Nb$_{2/3}$)$_{1-x}$Ti$_{x}$O$_{3}$ 
(PZN-x\%PT) oriented along an [001] direction, show exceptionally large piezoelectric deformations, more
than 1\% \cite{KUW,PARK}. In their pioneering work, Park and Shrout\cite
{PARK} proposed that the origin of the ultra-high strain values observed in
[001]-oriented PZN-8\%PT (8PT) was a rhombohedral-to-tetragonal phase transition
induced by the electric field. Later, Liu et al.\cite{LIU} reported similar
behavior in PZN-4.5\%PT (4.5PT) (see Fig. 1a). A revolution in the world of piezoelectric devices seems certain
to occur if the physical properties of such highly deformable materials can be understood and controlled.

Diffraction experiments on 8PT under an applied [001] field have
revealed the true long-range symmetry evolution to be from a rhombohedral to
a monoclinic phase \cite{NOH2}. These measurements were performed on
relatively thick samples, and a single tetragonal phase could not be reached
before a sample breakdown (see Fig. 1b). However it was shown that it
is the existence of such a monoclinic phase, rather than a tetragonal one,
which is crucial in explaining the outstanding properties of these materials.

Monoclinic phases have been observed in the temperature-composition phase
diagrams of three of the most important piezoelectric systems, Pb(Zr$_{1-x}$%
Ti$_{x})$O$_{3}$ (PZT) \cite{NOH1}, PZN-x\%PT \cite{COX,UES,LAO}, and very
recently also in Pb(Mg$_{1/3}$Nb$_{2/3}$)$_{1-x}$Ti$_{x}$O$_{3}$ (PMN-PT) 
\cite{XU,YE,KIA2,SIN1}, for compositions around the morphotropic phase boundary
(MPB), which represents the nearly vertical limit between the rhombohedral
and the tetragonal phases. Furthermore, it has been observed that the region of
stability of those new phases is enlarged following the application of an
electric field \cite{NOH2,GUO,OHW}. Optical \cite{PAIK} and x-ray diffraction
\cite{DUR2,VIE} measurements have also indicated a symmetry lowering in poled
8PT.

In contrast to the rhombohedral or tetragonal phases, the polarization
vectors in the monoclinic phases are no longer constrained to be directed
along a symmetry axis and can rotate within the monoclinic plane. However,
different polarization rotation paths have been observed in these materials,
resulting in two types of monoclinic distortion, M$_{A}$ (as found in PZT)
and M$_{C}$ (as found in 8PT) with space groups Cm and Pm,
respectively. As illustrated in Fig. 1c, in the M$_{A}$ type, the monoclinic
plane is the pseudo-cubic (1-10) plane and the unit cell is doubled in
volume with respect to the 4 \AA\ pseudo-cubic unit cell. In the M$_{C}$
type, the unit cell is primitive and the monoclinic plane is the
pseudo-cubic (010) plane. Very recently, Vanderbilt and Cohen \cite{VAN}
reported a natural derivation of these monoclinic phases by extending the
Devonshire expansion of the free energy up to eighth-order, which is
indicative of the large anharmonicity intrinsic to these
highly-piezoelectric materials. We are using their notation for the various
phases. A strong anharmonicity of the potential surfaces in lead oxides has
also been proposed by Kiat et al.\cite{KIA} based on experimental observations.

First-principles calculations by Bellaiche et al. have succeeded in
reproducing the monoclinic phase in PZT, both at zero electric field \cite
{BELL1} and also under an applied field \cite{BELL2}. Most importantly, the
calculations have shown that the huge increase in the piezoelectric
coefficients close to the MPB of these materials is directly related to
polarization rotation between the rhombohedral [111] and tetragonal [001]
polar axes\cite{FU} and, ultimately, to the existence of a monoclinic phase 
\cite{BELL1}. This has very important consequences for a fundamental
understanding of the piezoelectric properties in these compounds.

\begin{figure}[tbp]
\includegraphics[width=0.5\textwidth] {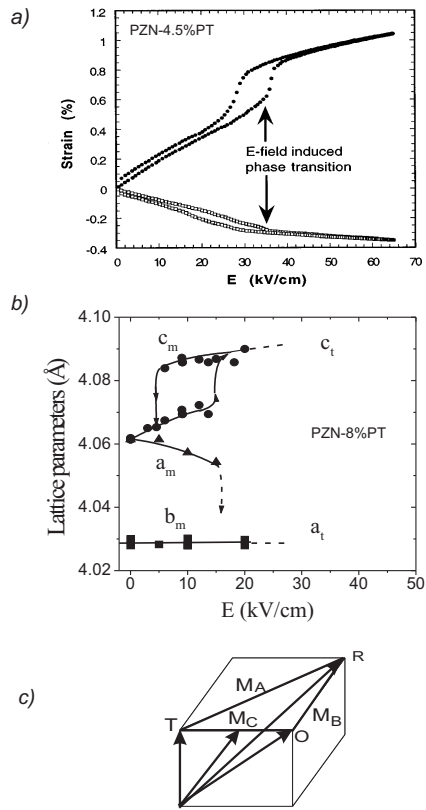}
\caption{ a) Macroscopic strain vs. electric field for 4.5PT after ref.
\onlinecite{LIU}. b) Lattice parameters vs. electric field for 8PT, 
adapted from ref.\onlinecite{NOH2}. c) Polarization vectors in the perovskite unit cell, shown by thick arrows.
The thick lines represent the paths followed by the end of the polarization
vector in the monoclinic phases, in between the rhombohedral (R), tetragonal
(T) and orthorhombic (O) phases. The M$_A$, M$_B$, M$_C$ notation is adopted
following Vanderbilt and Cohen\protect\cite{VAN}}
\end{figure}

In this new and revealing context, a detailed interpretation of the various
observations is required, with the goal of making the novel high-strain
piezoelectric systems as controllable and convenient as the PZT-based
materials in current use. We have accordingly performed high-energy x-ray
diffraction experiments on rhombohedral 4.5PT and 8PT crystals
with an electric field applied \textit{in-situ} along the [001] direction,
and studied how the symmetry evolves. A field-induced long-range tetragonal
phase has in fact been observed in both 4.5PT and 8PT
compositions, thus confirming Park and Shrout's thesis\cite{PARK}. These
experiments, together with those recently reported for 8PT \cite
{NOH2,VIE,OHW}, 9PT \cite{COX,UES} and xPT (x$\geq 10$\%) \cite
{LAO}, provide a much clearer picture of the field-induced behavior in
rhombohedral PZN-x\%PT. This work also provides the link between the
anomalous macroscopic strain values in these materials and the evolution of
the structural parameter under an applied field.

\section{Experimental}
4.5PT and 8PT samples were grown by the
high-temperature flux technique described in Ref. \cite{PARK}. The samples
were then oriented and cut in the form of rectangular parallelepipeds, with
sides ranging from 0.5 to 3mm, and at least two of the faces perpendicular
to an [001] direction. Gold was sputtered on two of the \{001\} faces of all
the samples, and thin wires were attached to enable an electric field to be
applied along the [001] direction. It should be noted that rhombohedral
PZN-x\%PT crystals as-grown have relaxor character and are, therefore,
crystallographically disordered. In order to induce the ferroelectric
long-range ordered state it is necessary to pole them under an electric
field. This poling is typically accomplished by applying a field of about 10
kV/cm at room temperature. If the electric field is applied along the [111] direction, no
change of symmetry is observed and the crystal in the ordered state is still
rhombohedral. As described above, in this work the crystals are poled along
the [001] direction. Very narrow mosaics were observed in the diffraction
patterns demonstrating the excellent quality of the crystals.

Single-crystal diffraction experiments were carried out at the National
Synchrotron Light Source, on beamlines X17B1 and X22A, with high-energy
x-rays of 67 keV ($\sim $0.18 \AA ) and 32 keV ( $\sim $0.38\ \AA ),
respectively. High-energy x-rays are needed to observe the crystal structure underneath the 
\textit{skin} of the sample, which extends a few microns below the surfaces\cite{NOH2,NOH3}and 
behaves differently from the bulk, as shown by Ohwada et al. \cite{OHW}. At X17B1, the high-flux monochromatic beam was obtained from a
superconducting wiggler device by use of a Si(220) crystal in Laue-Bragg
geometry. At X22A, the third-order reflection of a Si(111) monochromator crystal was
used to provide the 32 keV beam. Both beamlines are equipped with
four-circle Huber diffractometers, with Si(220) and Si(111)
analyzer-crystals mounted in the diffraction path at X17B1 and X22A,
respectively.

For the diffraction experiments with an \textit{in-situ }electric field
applied, a special sample holder was constructed in which the sample wires
were soldered to the high-voltage leads, and the samples were unclamped and
free to deform. The samples were coated with a silicone dielectric compound (%
\textit{GC Electronics-type Z5}) to prevent arcing, and kept in place with a
tiny dab of vacuum grease. The maximum value of the applied field was
limited by the dielectric breakdown of the samples; in the case of the
thinnest samples (d= 0.5mm), fields as high as 45 kV/cm were achieved.

\begin{figure}[tbp]
\includegraphics[width=0.5\textwidth] {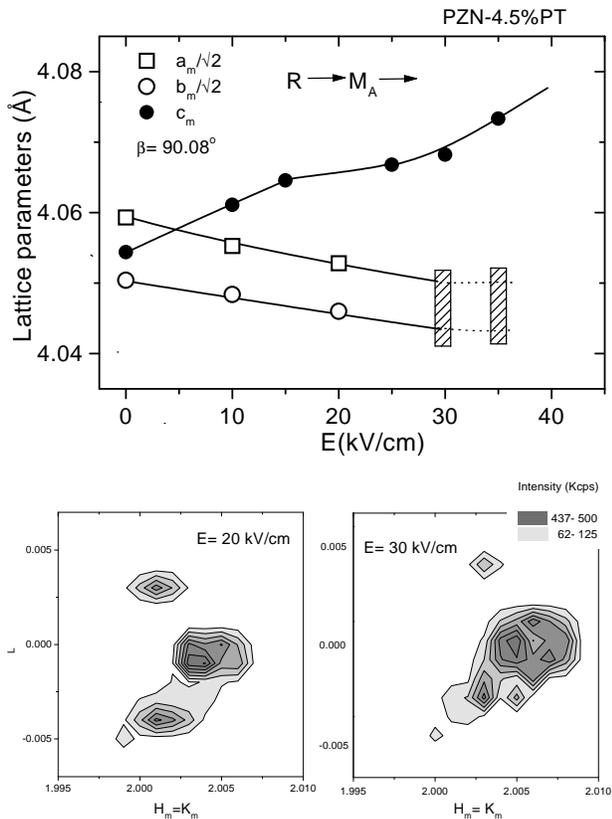}
\caption{Evolution of lattice parameters with an electric field applied
along the [001] direction in a 4.5PT crystal (dimensions 3x3x1mm$^3$) in the
rhombohedral and monoclinic phases, as observed by high-energy x-ray
diffraction(top). Mesh scans in the HHL zone of reciprocal space around
the pseudo-cubic (220) reflection are shown in the bottom plots for E= 20
kV/cm (bottom-left) and E= 30kV/cm (bottom-right). Intensities are on a linear scale}
\end{figure}

\section{Results}

\begin{figure}[tbp]

\includegraphics[width=0.5\textwidth] {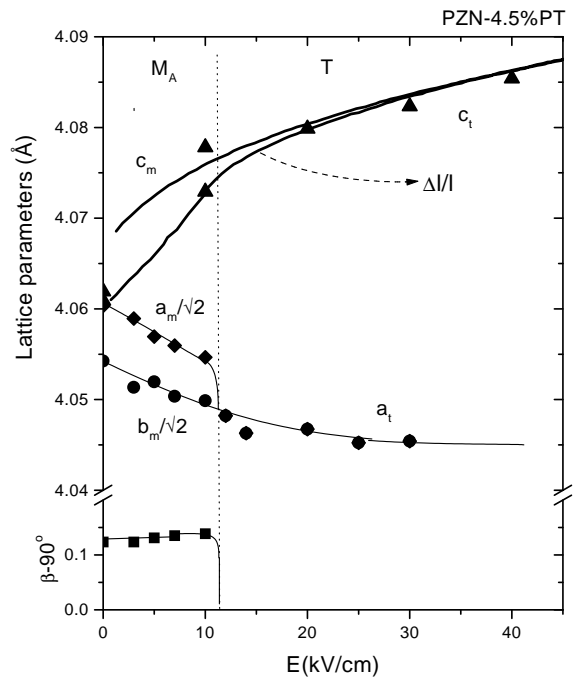}
\caption{Evolution of lattice parameters with an electric field applied
along the [001] direction in the second 4.5PT crystal in the monoclinic and
tetagonal phases, as observed by high-energy x-ray diffraction. The thick
lines represents the macroscopic unipolar strain along the [001] direction obtained
by dilatometric measurements on the same sample. The thinner lines are a
guide to the eye.}
\end{figure}

\begin{figure}[tbp]
\includegraphics[width=0.5\textwidth] {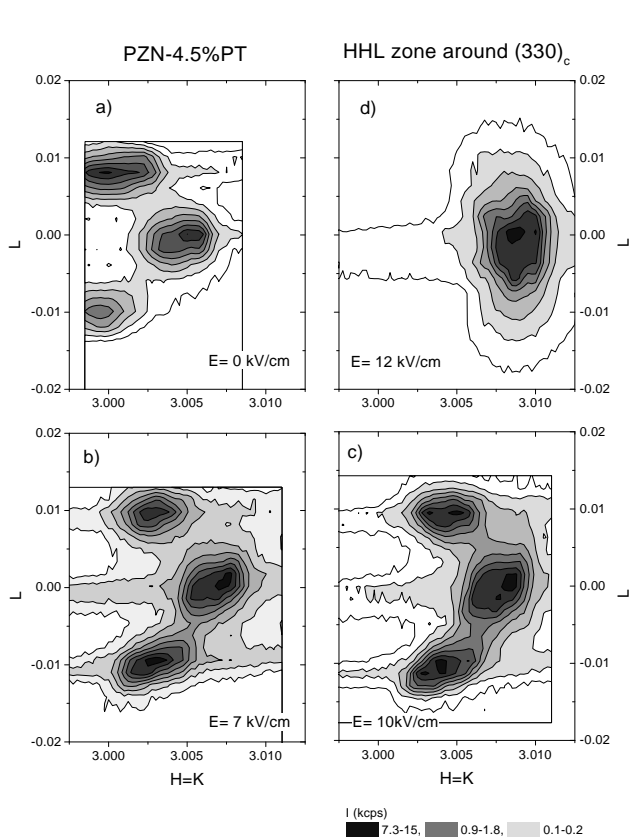}
\caption{Mesh scans around the pseudo-cubic (330) reflection in the HHL zone
of the reciprocal space for the second crystal of 4.5PT at E= 0 (a), 7 (b), 10 (c) and 12
(d) kV/cm. The intensities are plotted on a logarithmic scale}
\end{figure}

\begin{figure}[tbp]
\includegraphics[width=0.5\textwidth] {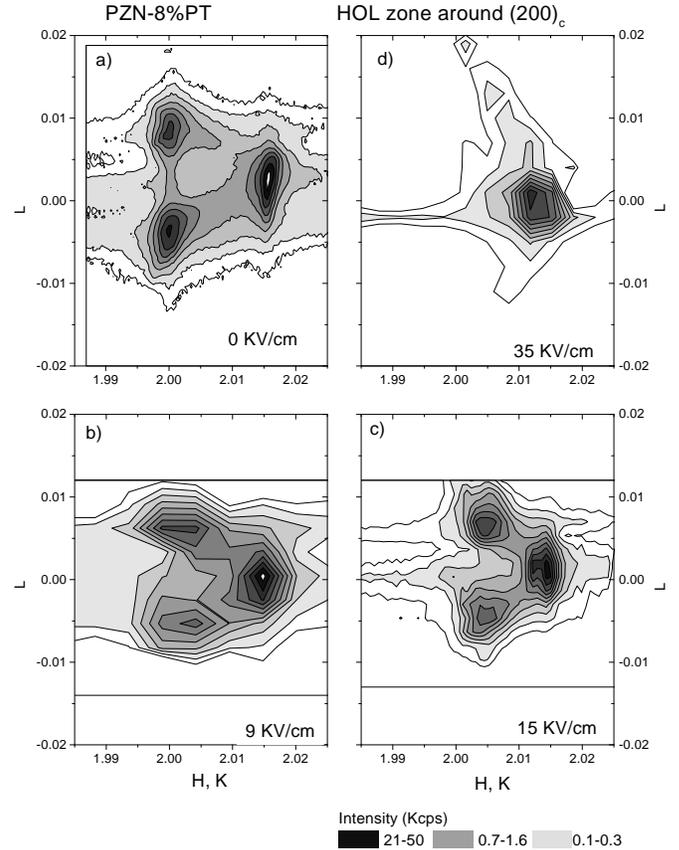} \vskip1pc
\caption{Mesh scans around the pseudo-cubic (200) reflection in the H0L (or
0KL) zone of reciprocal space for 8PT at E= 0 (a), 9 (b), 15 (c)
and 35 (d) kV/cm. The intensities are plotted on a logarithmic scale}
\end{figure}

Three 4.5PT crystals (A, B and C) were studied. In crystals A and B
the monoclinic phase was unambiguously observed for E < 30
kV/cm, consistent with the macroscopic strain measurements shown in Fig. 1a 
\cite{PARK}. The results obtained for crystal A are
summarized in Fig. 2. The initial state at E = 0, for which $a_{m}/\sqrt{2}$ 
$>c_{m}>$ $b_{m}/\sqrt{2}$, corresponds to a rhombohedral phase with $%
a_{r}=4.055$\AA\ and $\alpha _{r}=89.9^{o}$ \cite{note}. Under the application of an electric field along the [001]
direction, a monoclinic phase of M$_{A}$-type is induced for E $<$ 30 kV/cm, as shown by a steady decrease in $a_{m}/\sqrt{2}$ and $b_{m}/\sqrt{%
2}$, and a corresponding increase in $c_{m}$ (Fig. 2, top), and by a mesh
scan in the reciprocal HHL plane (the monoclinic plane) around the
pseudocubic (220) reflection at E = 20 kV/cm (Fig. 2, bottom-left). This
monoclinic phase is similar to the one observed in PZT\cite{NOH1}, 
in which $a_{m}$ and $b_{m}$ are rotated $45^{o}$ about the
pseudocubic [001] direction and are approximately equal to $a_{o}\sqrt{2}$,
and $c_{m}$ $\sim $ $a_{o}$, where $a_{o}$ is the pseudocubic lattice
parameter. The intensity distribution observed at 20 kV/cm in Fig. 2
(bottom-left) arises from the presence of four different M$_{A}$ domains
formed when the field is applied along the [001] direction, as previously
explained in ref. \onlinecite{YE}. Peaks are
observed at two different positions along the longitudinal scans around H =
K = 2, corresponding to $a_{m}$ and $b_{m}$ respectively. The splitting
between the pair of peaks at the smaller value of H = K (i.e. the larger
lattice parameter $a_{m}$) along the transverse (field) direction is related
to the monoclinic angle, $\beta =90.08^{o}$. For E = 30 kV/cm, the mesh
scans have a more complicated appearance in that most of the intensity has
coalesced within a single broad region, but with a distribution of lattice
parameters (Fig. 2, bottom-right), which we interpret as an incomplete
transformation to a tetragonal phase. However, even at E = 35 kV/cm, we did
not observe a single tetragonal phase, probably because of the existence of
several regions with different transformation fields.

The results obtained for crystal C were rather different; they showed
unambiguously a crossover between the monoclinic and tetragonal phases, but
at an anomalously low value of \symbol{126}11 kV/cm for this composition.
The evolution of the lattice parameters under a [001] electric field for
this crystal is shown in Fig. 3. In this case, the tetragonal phase,
characterized by two lattice parameters $a_{t}$ and $c_{t}$, is indeed
observed at high fields, as proposed by Park and Shrout on the basis of
macroscopic measurements \cite{PARK}. When the field
is decreased, the tetragonal distortion $c_{t}/a_{t}$ also decreases;
however, the tetragonal phase does not transform directly into a
rhombohedral phase at low fields. Instead, a monoclinic phase is observed as 
$a_{t}$ splits into $a_{m}$ and $b_{m}$, and the angle $\beta $ between $%
a_{m}$ and $c_{m}$ becomes slightly greater than $90^{o}$. This monoclinic
phase is also of M$_{A}$-type, and as can be seen in Fig. 3, the crystal
remains monoclinic even at E = 0. It is noteworthy that at this point, $%
c_{m}\sim a_{m}/\sqrt{2}$, corresponding to the singular case of a monoclinic
cell with the polarization vector lying along the rhombohedral polar axis
[111], probably as a result of the underlying monoclinic distortion of the
oxygen octahedra. In order to recover the initial rhombohedral phase, it
would be necessary to apply a field in the reverse direction.

Fig. 4 shows the region of the reciprocal HHL plane around the pseudocubic
(330) reflection for several different values of E. The intensity
distributions observed in Figs. 4a-c (E = 0, 7 and 10 kV/cm, respectively)
arise from the four different M$_{A}$ monoclinic domains formed when the
field is applied along the [001] direction, as noted previously. Once again,
peaks are observed at two different positions along the longitudinal scans
around H = K = 3, corresponding to $a_{m}$ and $b_{m}$ respectively. Note
that H = K = L = 3 is defined in order to reflect the value of $c_{m}$ at E
= 0, so that the value H = K = 3 in the scans means that $a_{m}$ = $c_{m}$.
The splitting between the pair of peaks at the smaller value of H = K
corresponds to a monoclinic angle of $\beta $ = 90.13$^{o}$ . As the field
is increased, both $a_{m}$ and $b_{m}$ decrease, as shown by the shifts to
larger H = K values. At E = 12 kV/cm, the intensity has coalesced into a
single peak (Fig. 4d) and the crystal is in the tetragonal phase (i.e. $a_{m}
$ $=$ $b_{m}$, $\beta $ $=90^{o}$). From Fig. 3, it is seen that the
transformation occurs at about 11 kV/cm. The composition of the crystal C was confirmed 
using dielectric 
measurement, with T$_{max}$ being $~170^{o}$C. In addition to the exceptionally 
low phase transition field, the strain behavior associated with the phase transition in Fig. 3 
does not show the typical hysteretic jump at the transition field. We suspect that 
the defect structure and subsequent abnormal domain configuration may affect its phase 
transition behavior. Further investigation is on going.

4.5PT therefore behaves differently than 8PT, as can be readily inferred from Fig.1b. As
reported in ref. \onlinecite{NOH2}, application of a [001] electric field to the latter yields a monoclinic 
cell of M$_{C}$-type, with a primitive cell of about 4 \AA\ along the edge. In the 8PT case the
characteristic intensity distribution resulting from the four possible monoclinic domains is
observed in the 0KL (or H0L) zone of reciprocal space, since the
monoclinic angle is now contained within the (010) plane. Figs. 5a-d show such
intensity distributions around the pseudo-cubic (200) reflection, for several different 
values of the applied electric field, E= 0, 9, 15 and 35 kV/cm,
respectively. Unlike 4.5PT, $b_{m}$ remains approximately constant at K= 2.015 as the field is increased, 
while $a_{m}$ decreases,
approaching $b_{m}$ (see also Fig.1b). At sufficiently high fields, $a_{m}$
becomes equal to $b_{m}$ and the crystal becomes tetragonal, showing a single
reflection around the (200) point in reciprocal space (Fig.5d), as also
illustrated by the dotted lines in Fig. 1b.

\begin{figure}[tbp]
\includegraphics[width=0.5\textwidth] {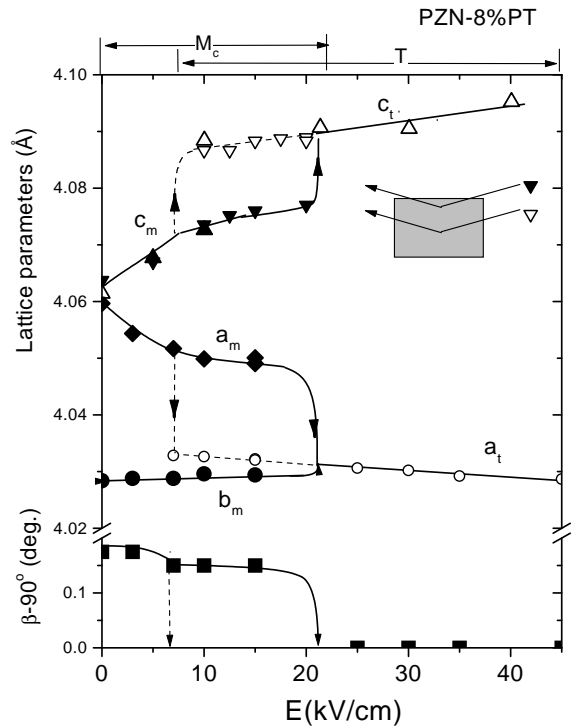}
\caption{Evolution of lattice parameters with an electric field applied along the [001] direction for a 8PT
crystal. Solid symbols represent the monoclinic lattice parameters. Open
symbols represent the tetragonal lattice parameters. Inverted triangles show the
results of measurements of the $c$ parameter with a 100 $\protect\mu $m
beam at two different depths in the crystal , as illustrated schematically in the inset.}
\end{figure}

When the field is removed, the crystal does not become rhombohedral, but 
resembles 4.5PT in this respect, with $a_{m}$ =$c_{m}\neq b_{m},$ $\beta \neq
90^{o}$. However, unlike M$_{A}$, in this case the equality $a_{m}$ =$c_{m}$ corresponds to a higher 
symmetry than M$_{C}$, namely orthorhombic, O (see also Fig. 1c)\cite{NOH1}. 
This "pseudo-monoclinic" O phase (O*) has recently been observed in
the phase diagram of PZN-x\%PT for 9PT and 10PT\cite{COX,UES,LAO}.

As we have mentioned, the crossover between the monoclinic and tetragonal
phases depends on the sample composition, as well as on the experimental
conditions, such as mechanical clamping. High energy x-ray diffraction has revealed that 
the critical fields at which the monoclinic-to-tetragonal phase change takes
place are different across the sample thickness\cite{OHW}, most likely due
to a distribution of strain. Fig. 6 shows the lattice parameters of one of
the 8PT crystals under an [001] electric field. The results are very
similar to those in Fig. 1b\cite{NOH2} but in this case the tetragonal phase
could be reached. We observed an intermediate region (7$<$ E $<$ 20 kV/cm) in which the two phases, monoclinic $M_{C}$ and
tetragonal, coexist. Measurements of the $c$ lattice parameter made with a
narrow beam about 100 $\mu $m in width (inverted triangles in Fig. 6) showed that the
tetragonal phase was indeed reached at different field values across the sample thickness, between
7-20 kV/cm.

\section{Discussion}

The most outstanding feature of the [001]-oriented rhombohedral
PZN-x\%PT crystals is the extremely high deformation that can be obtained under
the application of an electric field. Of paramount interest from the applications 
point-of-view is that large elongations can be obtained with
very little hysteresis, which means virtually no domain-wall motion. As
observed in the $\Delta l/l$ measurements plotted in Fig.3, the macroscopic
deformation of about 0.6\% in 4.5PT for E$\leq 40kV/cm$, corresponds 
very closely to the elongation of the unit cell along the field direction.
The observed total strain is due partly to polarization rotation and
elongation in the M$_{A}$ phase between [111] and [001] and partly due to
the $c$-axis elongation of the unit cell in the tetragonal phase. However, a
change of slope is clearly observed in both $\Delta l/l$ and the $c$ parameter with applied
field at the M$_{A}$-T phase transition, from which it can be inferred directly 
that the piezoelectric modulus (elongation per volt) in the
monoclinic phase is about five times larger than that in the tetragonal phase.

The observed macroscopic strain depends on the initial domain configuration,
and thus, on the previous history of the crystal. The coincidence of the
microscopic and macroscopic measurements in Fig. 2 indicates that the
crystal consisted only of domains with polarizations $35^{o}$ away from the
field direction [001] (see refs. \onlinecite{PARK,CRO}). When the field is
applied, the polarization of the domains rotates until the crystal becomes
tetragonal and monodomain with the polarization parallel to the field.

However, a negative electric field or mechanical pressure along the field
direction can favor domains with polarizations closer to the plane
perpendicular to the field, as recently described for the PZT case\cite{CHA}%
. This fact can explain the dramatic differences observed between the
stress-free-bipolar and slightly-clamped-unipolar strain curves reported by
Viehland \cite{VIE} on 8PT crystals. The deep strain level of -0.6\%
and the aggregate strain level of 1.2\% observed in the bipolar strain curve
of <001>-oriented 8PT crystals correspond to
the maximum spontaneous strain observed for $M_{C}$, i.e., the strain between the
monoclinic $a$ and $b$, and $b$ and $c$ axes, respectively (see Fig. 1b).

\begin{figure}[tbp]
\includegraphics[width=0.5\textwidth] {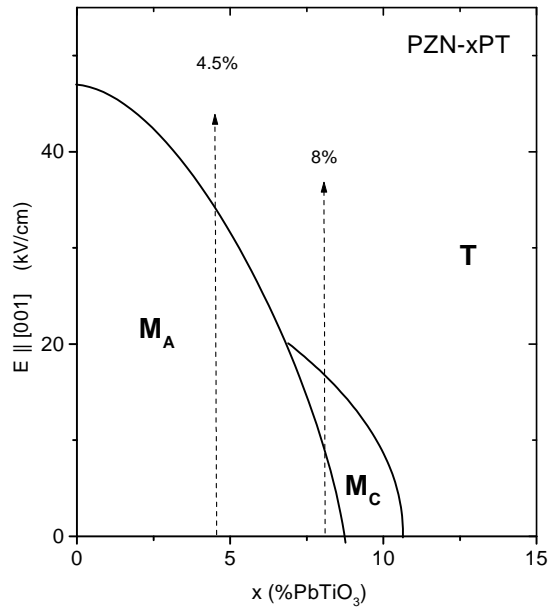}
\caption{ Sketch of the [001]-electric field vs. composition phase diagram
for PZN-x\%PT}
\end{figure}

The extreme sensitivity of these materials to the experimental conditions,
in particular to stress, is an indication of the near-degeneracy of the
different phases around their MPB's. Although the exact field values at
which the phase transitions happen would need to be determined for very
specific experimental conditions, we still can sketch an electric field vs.
composition phase diagram for rhombohedral PZN-x\%PT as shown in Fig. 7.
Close to the MPB at 8PT, the M$_{A}$ phase is only stable at very
low fields\cite{OHW}, while no M$_{C}$ phase has been so far observed for
4.5PT.

\begin{figure}[tbp]
\includegraphics[width=0.5\textwidth] {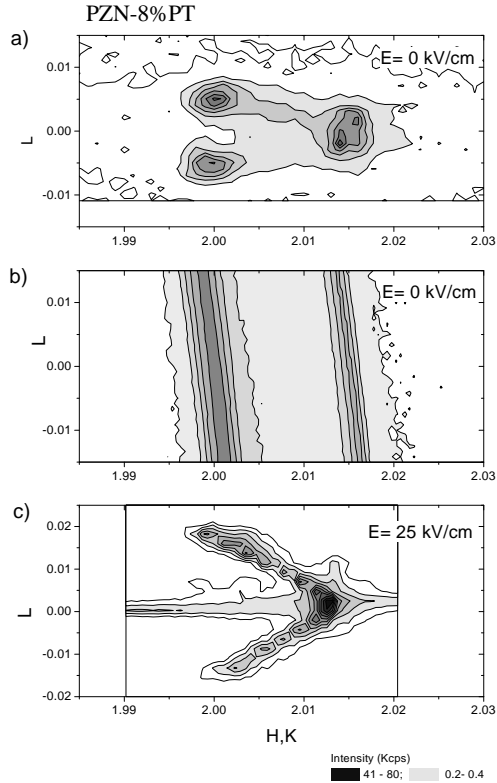}
\caption{a) Typical profiles observed for 8PT in the HOL zone of reciprocal space 
around the pseudo-cubic (200) reflection at E= 0. b) Example of an unusual profile observed in the same region. c)
An unusual intensity distribution observed for the same composition in the
tetragonal phase. The intensities are on a logarithmic scale}
\end{figure}

An intricate and history-dependent domain configuration is another
consequence of the high degeneracy of the various phases displayed by the
PZN-xPT system, which can also produce some exotic and unusual diffraction
intensity distributions, especially for compositions very close to the MPB.
Fig. 8a shows the three-peak pattern usually observed in 8PT
crystals in the HOL zone around the pseudo-cubic (200) reflection at E=0, arising 
from the four different monoclinic domains\cite{NOH2}. However, on one occasion a very 
interesting rod-like intensity distribution was observed
in the same crystal upon the removal of the electric field (Fig. 8b), the
d-spacings of the rods being identical to those of the Bragg peaks.
Unfortunately, we were never able to reproduce such a rod pattern, but
we speculate that this interesting behavior may be due to a complicated
domain formation. Another example of unusual diffuse scattering is shown in
Fig. 8c., corresponding to the same region of the reciprocal space of a 8PT crsytal under 
a 25 kV/cm [001]-electric field. 
Most of the crystal has transformed into the tetragonal phase characterized
by a single lattice parameter in this zone (as in Fig.5d). The sharpness of
this reflection shows the excellent quality of the crystal and the accuracy
of the electric field orientation. However, when plotted on a logarithmic
scale (as in Fig. 8c), a fascinating fish-like shape is revealed, indicating
that a small fraction of the sample retains a distribution of both,
lattice parameters and monoclinic angle.

To conclude, the polarization rotation path has been investigated for
rhombohedral 4.5PT and 8PT single crystals under a [001]
electric field. In 4.5PT the polarization vector rotates directly from
[111] towards [001] via a monoclinic M$_{A}$ phase. In 8PT, which lies
closer to the MPB, the polarization vector jumps at a relatively low field to
the (010) plane and rotates in this plane, via a monoclinic M$_{C}$
phase, towards [001] when the field is increased\cite{NOH2}. For both
compositions a single tetragonal phase has been observed at high fields. The
behavior of the lattice parameters as a function of electric field can account for
the ultra-high macroscopic piezoelectric deformations in terms of the
microscopic deformation of the unit cell (rotation plus elongation).

On occasions, unique contour plots have been recorded, especially for the
8PT crystals, which show very peculiar intensity distributions and are
believed to reflect the existence of heavily twinned materials and complicated local effects.
Further work is needed to fully understand some of these features; in
particular, a detailed study of the complicated diffuse scattering would
 provide very useful information about the local order in these materials.
However, we may safely conclude that all the reported observations are
consequences of the high anharmonicity and the delicate energy balance
between the different phases in these highly deformable materials.

\begin{acknowledgements}

Stimulating discussions with G. Baldinozzi, L. Bellaiche, L.E. Cross, B. Dkhil, M. Durbin, K. Hirota, K. Ohwada, J-M.
Kiat, D. Vanderbilt, D. Viehland, and T. Vogt, as well as the technical
support of A. Langhorn are gratefully acknowledged. Financial support by DOE
under contract No. DE-AC02-98CH10886 and the Office of Naval Research is also acknowledged.

\end{acknowledgements}


\begin{references}
\bibitem{KUW}  J. Kuwata, K. Uchino, and S. Nomura, Jpn.J. Appl. Phys., Part1 {\bf 21}, 1298 (1982).

\bibitem{PARK}  S-E. Park and T.R. Shrout, J. Appl. Phys. {\bf 82}, 1804 (1997).

\bibitem{LIU}  S-F. Liu, S-E. Park, T. Shrout and L.E. Cross, J. Appl. Phys.
{\bf 85}, 2810 (1999)

\bibitem{NOH2}  B. Noheda, D.E. Cox, G. Shirane, S-E.
Park, L.E. Cross, and Z. Zhong, Phys. Rev. Lett. {\bf 86}, 3891 (2001).


\bibitem{NOH1}  B. Noheda, D. E. Cox, G. Shirane, J.A. Gonzalo, L.E. Cross, and S-E. Park, 
Appl. Phys. Lett {\bf 74}, 2059 (1999); B. Noheda, J.A. Gonzalo, L.E. Cross, R. Guo, S-E. Park, D.E. Cox, and G. Shirane, 
Phys. Rev. B {\bf 61}, 8687 (2000); B. Noheda, D.E. Cox, G. Shirane R, Guo, B. Jones, and L.E. Cross , Phys. Rev. B. {\bf 63}, 14103 (2001).

\bibitem{COX}  D.E. Cox, B. Noheda, G. Shirane, Y. Uesu, K. Fujishiro,
and Y. Yamada, Appl. Phys. Lett {\bf 79}, 400 (2001).


\bibitem{UES}  Y. Uesu, M. Matsuda, Y. Yamada, K. Fujishiro, D.E. Cox, B. Noheda and G. Shirane, 
J. Phys. Soc. Japan (in press). E-print: cond-mat/00106552.


\bibitem{LAO}  D. La-Orauttapong, B. Noheda, Z-G. Ye, P.M. Gehring, J. Toulouse, D.E. Cox, and G. Shirane, Phys. Rev. B (in press). 
E-print: cond-mat/0108264

\bibitem{XU}  G. Xu, H. Luo, H. Xu and Z. Yin, Phys. Rev. B {\bf 64}, 020102(2001)

\bibitem{YE}  Z-G. Ye, B. Noheda, M. Dong, D.E. Cox, and G. Shirane, Phys. Rev. B {\bf 64}, 184114 (2001). 

\bibitem{KIA2} J.-M. Kiat, Y. Uesu, B. Dkhil, M. Matsuda, C. Malibert, and G. Calvarin, Phys. Rev. B. (in press). 

\bibitem{SIN1} A. K. Singh and D. Pandey, J. Phys.:Condens. Matt. {\bf 13}, L931 (2001)
\bibitem{GUO}  R. Guo, L.E. Cross, S-E. Park, B. Noheda,
D.E. Cox , and G. Shirane, Phys. Rev. Lett. {\bf 84}, 5423 (2000).


\bibitem{OHW}  K. Ohwada, K. Hirota, P.W. Rehrig, P.M. Gehring, B.
Noheda, Y. Fujii, S-E. Park, and G. Shirane, J. Phys. Soc. Japan {\bf 70}, 2778 (2001)


\bibitem{PAIK}  D-S. Paik, S-E. Park, S. Wada, S-F. Liu and T. Shrout, J.
Appl. Phys. {\bf 85}, 1080 (1999)

\bibitem{DUR2}   M.K. Durbin, J.C. Hicks, S.-E. park, and T.R. Shrout, J. Appl. Phys. {\bf 87}, 8159 (2000).

\bibitem{VIE}  D. Viehland, J. Appl. Phys. {\bf 88}, 4794 (2000); D. Viehland, J.
Powers and L. Ewart, J. Appl. Phys. {\bf 88}, 4907 (2000).


\bibitem{VAN}  D. Vanderbilt and M.H. Cohen, Phys. Rev. B. {\bf 63}, 94108 (2001).


\bibitem{KIA}  J-M. Kiat, G. Baldinozzi, M. Dunlop, C. Malibert, B. Dkhil,
C. Menoret, O. Masson, and M.T. Fernandez-Diaz, J. Phys.: Condens. Matter 
{\bf 12}, 8411 (2000)

\bibitem{BELL1}   L. Bellaiche, A. Garcia, and D. Vanderbilt, Phys. Rev. Lett. {\bf 84}, 5427 (2000)


\bibitem{BELL2}  L. Bellaiche, A. Garcia, and D. Vanderbilt, Phys. Rev. B {\bf 64}, 060103(R) (2001)


\bibitem{FU}  H. Fu and R.E. Cohen, Nature {\bf 403}, 281
(2000).



\bibitem{NOH3} B. Noheda, D.E. Cox, and G. Shirane, Ferroelectrics (in press). E-print: cond-mat/0109545.

\bibitem{note} It can be easily shown that in the rhombohedral phase $a_m$, $b_m$ and $c_m$ are not longer independent, 
and are related to the rhombohedral lattice parameters, $a_{r}$ and $\alpha_{r}$, as follows: 
$a_m$= $2a_{r}cos(\alpha _{r}/2)$, $b_m$= $2a_{r}sin(\alpha _{r}/2)$ and $c_{m}$=$a_{r}$. 
The monoclinic angle, $\beta $, is then defined as $180^{o}$-$\phi $, where $cos(\phi )=[1-2sin^{2}(\alpha _{r}/2)]/
cos(\alpha _{r}/2)$.
\bibitem{CRO} L. E. Cross, AIP Conference Proceedings, vol. 535, pp.1-15 (2000)
\bibitem{CHA}  P. Chaplya and G. P. Carman, J. Appl. Phys. {\bf 90}, 2578 (2001)
\end{references}
\end{document}